\documentclass{emulateapj}

\shorttitle{Secular Gravitational Instability of the Dust Layer}
\shortauthors{Takeuchi \& Ida}
\slugcomment{Accepted by the Astrophysical Journal}

\begin{document}

\title{Minimum Dust Abundances for Planetesimal Formation via Secular
Gravitational Instabilities}
\author{Taku Takeuchi\altaffilmark{1} and Shigeru Ida}
\affil{Department of Earth and Planetary Sciences, Tokyo Institute of Technology, Meguro-ku, Tokyo, 152-8551, Japan}

\begin{abstract}
We estimate minimum dust abundances required for secular gravitational
instability (SGI) to operate at the midplane dust layer of protoplanetary
disks. For SGI to be a viable process, the growth time of the instability
$T_{\rm grow}$ must be shorter than the radial drift time of the dust
$T_{\rm drift}$. The growth time depends on the turbulent diffusion parameter
$\alpha$, because the modes with short wavelengths are stabilized by turbulent
diffusion. Assuming that turbulence is excited via the
  Kelvin-Helmholtz or streaming instabilities in the dust layer, and
that its strength is controlled by the energy supply rate from dust
accretion, we estimate the diffusion parameter 
and the growth time of the instability. The condition $T_{\rm grow}
<T_{\rm drift}$ requires that the dust abundance must be greater than a critical
abundance $Z_{\rm min}$, which is a function of the Toomre parameter
$Q_{g}$ and aspect ratio $h_{g}/r$ of the gas disk. For a wide 
range of parameter space, the required dust abundance is less than $0.1$. A
slight increase in dust abundance opens a possible route for the dust to
directly collapse to planetesimals.

\end{abstract}


\altaffiltext{1}{taku@geo.titech.ac.jp}

\setcounter{footnote}{2}

\section{Introduction}

Formation processes of planetesimals have not been well understood in planet
formation theory. The gravitational instability of the dust layer at the
midplane of a protoplanetary disk has been proposed as a possible route
to planetesimal formation. The classical scenario of gravitational instability
requires high densities of dust layers to surpass the Roche limit
(Goldreich \& Ward 1973; Sekiya 1983). 
Dust sedimentation only in the vertical direction hardly achieves
such a high density (Sekiya 1998). The dust layer becomes turbulent
via Kelvin-Helmholtz (KH) or streaming instabilities when its dust density
exceeds the gas density, which is much less than the Roche density, and
further dust settling is significantly suppressed (Chiang \& Youdin 2010
for review). Radial drift of the dust caused by gas drag may provide a
possible route for further accumulation of the dust. When the
dissipative effects of gas drag are included, the dust layer is
secularly gravitationally unstable to the modes of dust accumulation in
the radial direction (Ward 2000; Youdin 2005a, 2005b). This secular
gravitational instability (SGI) occurs in any dust layers even if their
densities are small (i.e., there is no criterion on their Toomre's $Q$
values for SGI; Youdin 2011; Shariff \& Cuzzi 2011; Michikoshi et
al. 2012; these papers are referred hereafter as Y11, SC11, and MKI12,
respectively).

However, the growth rate of SGI is greatly suppressed by gas drag,
particularly for small particles, and could be too slow to have any
effect on planetesimal formation. Y11, SC11, and MKI12 argue that the growth
timescale must be shorter than the lifetime of the dust drifting toward 
the star owing to gas drag. The wavelength of the most unstable mode is
determined by the balance between the radial accumulation of dust
particles due to self-gravity and their diffusion due to gas
turbulence. Only perturbations 
with wavelengths sufficiently long are unstable. For stronger turbulence,
wavelengths of the unstable modes, as well as growth timescales, are
longer. Y11 and SC11 have calculated the growth timescales for various values
of the turbulent diffusion parameter $\alpha$, and have derived an upper
limit on 
$\alpha$ required for the growth timescale to be shorter than the drift
timescale. Their approach is general and can be applied for any disk
turbulence, provided that the value of $\alpha$ is known.

In this paper, we focus on disk turbulence induced in the dust layer, assuming
that the gas disk itself is a globally laminar flow. In such a disk, the
concentration of dust particles at its midplane triggers
turbulence. The velocity difference between the dust and gas induces the
KH or streaming instabilities and causes turbulence in the dust layer (Chiang \& Youdin 2010). 
The ultimate energy source for turbulence
is the accretion energy of the dust drifting toward the star, regardless
of the kind of instability that occurs in the dust layer. Thus, a simple
energetics can be applied to estimate the turbulence strength, provided
that the dust accretion velocity is known. Takeuchi et al. (2012,
hereafter T12) calculate the energy supply rate to turbulence, using the
classical formulae on the drift velocity of the dust (Nakagawa et
al. 1986; Weidenschilling 2003; Youdin \& Chiang 2004), and then
estimate the turbulence strength or the value of $\alpha$. T12 has shown
that, from the comparison between the estimated value of $\alpha$ and
the recent results of numerical simulations of turbulence by Johansen et
al. (2006) and Bai \& Stone (2010), the turbulence strength excited via
KH and/or streaming instabilities can be estimated from the dust
accretion rate.

While Y11, SC11, and MKI12 have derived the growth time of SGI as a function of
$\alpha$ and other disk parameters (Equation (51) of Y11), T12 has obtained
$\alpha$ values of turbulence in the dust layer (Equation
(\ref{eq:alpha}) below).
Combining these results gives an expression of the growth time without
the unknown parameter $\alpha$, as shown in Section \ref{sec:growthtime}. Consequently, in
Section \ref{sec:comptime}, the criterion for SGI to operate is obtained as a condition
on the dust abundance $Z$ in the disk. For SGI to operate faster than the dust
drift timescale, the dust abundance must be greater than a specific critical
value $Z_{\rm min}$, which is a function of the Toomre $Q$ value of the gas
disk and the deviation fraction $\eta$ of the gas velocity from the Keplerian
velocity. The $Z_{\rm min}$ value derived in this paper gives the minimum dust
abundance required for SGI, because we consider only turbulence induced in the
dust layer. If turbulence were caused by another mechanism such as
magneto-rotational instability (MRI) of the gas disk, $\alpha$ would be larger
than the value we adopted, and consequently the required value of
$Z_{\rm min}$ would increase. Hence, $Z_{\rm min}$ derived in this paper
is considered as the minimum value required for SGI in realistic disks.

\section{Condition for Secular Gravitational Instability}

\label{sec:tgrow}

We consider a protoplanetary disk initially in a laminar flow state.
Sedimentation of dust particles to the midplane of the disk induces
turbulence via KH or streaming instabilities caused by velocity
differences between the 
dust and gas. A (quasi-)steady dust layer forms when turbulent diffusion
matches dust settling, in which steady state turbulence is maintained. We
neglect intermittency of turbulence. If an extra source of turbulence is
present such as global turbulence of the gas disk via MRI, the
turbulence would be stronger than that 
excited only by the dust-gas velocity difference. Thus, in this paper, we
consider the minimum strength of turbulence. 

The dust particles are characterized by their stopping time $t_{s}$,
which is the timescale of damping the velocity difference from the
gas. We define the non-dimensional stopping time, $T_{s}=t_{s}
\Omega_{\rm K}$, normalized by the Keplerian frequency $\Omega_{\rm K}$.
In this study, we focus on the dynamics of relatively small
particles such that $T_{s}\la1$. Particles with $T_{s}\sim1$ experience
high speed collisions and the fastest radial drift, which likely hinder
particle growth. SGI of the dust layer is a possible route for directly
forming large bodies ($T_{s}\gg1$) from small particles ($T_{s}\ll1$).

In the following subsections, we compare the growth time of SGI with the orbital
drift time of the dust particles due to gas drag in order to determine the
condition for SGI to be a relevant process for planetesimal formation.

\subsection{Timescale of Secular Gravitational Instability}

\label{sec:growthtime}

In his Equation (51), Y11 shows that the growth time of SGI for small
particles with $T_{s}\la1$, normalized by the Keplerian time $\Omega_{\rm K}^{-1}$, is
\begin{equation}
T_{\rm grow}\approx\frac{\alpha Q_{g}^{2}}{Z^{2}T_{s}^{2}} \ .
\label{eq:tgrow}
\end{equation}
Here $\alpha$ is the turbulent diffusion parameter (see below), $Q_{g}$ is
the Toomre stability parameter of the gas disk,
\begin{equation}
Q_{g}=\frac{c_{g}\Omega_{\rm K}}{\pi G\Sigma_{g}}~,
\end{equation}
where $c_{g}$ is the sound speed of the gas, $G$ is the gravitational constant,
and $\Sigma_{g}$ is the surface density of the gas disk. The disk
``metallicity'' $Z$ is the ratio of the surface densities between the
dust and gas, $Z=\Sigma_{d}/\Sigma_{g}$. The growth time depends on the
turbulent diffusion parameter $\alpha$ through its stabilizing effect for perturbations with short wavelengths.

Turbulence in the dust layer is induced by velocity differences between the
dust and gas. If the drag force between the gas and dust were not
effective, the dust would orbit with the Keplerian velocity $v_{\rm K}$, while the
gas would orbit with a sub-Keplerian velocity
\begin{equation}
v_{g}=(1-\eta)v_{\rm K}~,
\end{equation}
where $\eta=-(2\rho_{g}r\Omega_{\rm K}^{2})^{-1}\partial P/\partial r\sim
(c_{g}/v_{\rm K})^{2}\sim10^{-3}-10^{-2}$, $\rho_{g}$ is the gas density, and $P$
is the gas pressure. In the calculation of $\alpha$, a slightly different
definition of $\eta$ is useful (see T12 or Appendix \ref{sec:app-approx}
in this paper). We introduce $\tilde{\eta}$ defined by
\begin{equation}
\tilde{\eta}=\left(  \frac{v_{\rm K}}{c_{g}}\right)  ^{2}\eta^{2}=C_{\eta}
\eta \ ,
\end{equation}
where $C_{\eta}$ is a factor of the order of unity, which depends on the radial
profile of the gas disk. In this paper, we adopt $C_{\eta}=1$ (i.e.,
$\tilde{\eta}=\eta$) for simplicity, except in Section \ref{sec:diskmodel} where
specific disk models are considered. 

T12 shows that the dust particles accrete toward the star due to
  gas drag and supply energy to turbulence via KH and/or streaming
  instabilities. Dust accretion is either caused by the gas drag acting
on individual particles or 
by turbulent drag acting on the surface of the dust layer. The former
(individual drag) is effective if the dust-to-gas ratio at the disk midplane,
$f_{\rm mid}=\left.  \rho_{d}/\rho_{g}\right\vert _{z=0}$, is less than unity, and
the latter (collective drag) is effective if $f_{\rm mid}\ga1$. Appendix
\ref{sec:app-approx} briefly summarizes dust accretion due to both drag
types (see T12 for detailed discussions). The turbulence strength or the
parameter $\alpha$ is determined by the energy supply rate from dust accretion
toward the star. The approximate expression of $\alpha$ for $T_{s}\la1$
particles is given by
\begin{equation}
\alpha\approx\left[  \left(  C_{1}C_{\rm eff}\tilde{\eta}Z\right)  ^{-\frac{2}{3}
}+\left(  C_{2}C_{\rm eff}\tilde{\eta}Z^{-1}\right)  ^{-2}\right]  ^{-1}
T_{s}~,\label{eq:alpha}
\end{equation}
where $C_{\rm eff}=0.19$ is the energy supply efficiency (see T12), and
$C_{1}=1.0$ and $C_{2}=1.6$ are the numerical factors. This expression
connects the approximate formulae of $\alpha$ for the two limiting
regimes of $f_{\rm mid}\ll1$ and $f_{\rm mid}\gg1$ (Equations
(\ref{eq:aplha-apx1}) and (\ref{eq:aplha-apx2})). 
Note that the condition $f_{\rm mid}\ll1$ (or  $f_{\rm mid}\gg1$) corresponds to the
condition $Z\ll(C_{\rm eff}\tilde{\eta})^{1/2}$ (or $Z\gg(C_{\rm eff}\tilde
{\eta})^{1/2}$) (see Equation (\ref{eq:fmid2})). The numerical factors $C_{1}$ and
$C_{2}$ are adjusted to make an appropriate fit to the overall behavior of the
numerical result. A comparison of this approximate expression to the
numerically calculated $\alpha$ is shown in Appendix \ref{sec:app-approx}.
Substituting Equation (\ref{eq:alpha}) in Equation (\ref{eq:tgrow}) gives
\begin{equation}
T_{\rm grow}\approx\frac{Q_{g}^{2}}{Z^{2}T_{s}}\left[
\left(  C_{1}C_{\rm eff}\tilde{\eta}Z\right)^{-\frac{2}{3}}+\left( C_{2}
C_{\rm eff}\tilde{\eta}Z^{-1}\right)^{-2}\right]^{-1}~,
\label{eq:tgrow2}
\end{equation}
showing that $T_{\rm grow}$ is a function of the disk
parameters $(\tilde{\eta},Z,Q_{g})$ and is inversely proportional to $T_{s}$.

\subsection{Timescale of Radial Drift}

According to T12, the non-dimensional timescale of the radial drift of dust
particles due to gas drag is estimated as
\begin{equation}
T_{\rm drift}\approx\frac{f_{\rm mid}+1}{2\eta T_{s}}~,
\label{eq:Tdrift}
\end{equation}
where the midplane dust-to-gas ratio $f_{\rm mid}$ is approximately given
by
\begin{equation}
f_{\rm mid}\approx\left(  \frac{Z^{2}}{C_{1}C_{\rm eff}\tilde{\eta}}\right)
^{\frac{1}{3}}+\left(  \frac{Z^{2}}{C_{2}C_{\rm eff}\tilde{\eta}}\right)
~.\label{eq:fmid}
\end{equation}
In the above estimate, both individual and collective drags are considered.
Derivation of the above expressions are described in Appendix
\ref{sec:app-approx}. The timescale of the radial drift is a function of
$(\tilde{\eta},Z)$ and is inversely proportional to $T_{s}$.

\begin{figure}
\epsscale{1.15} 
\plotone{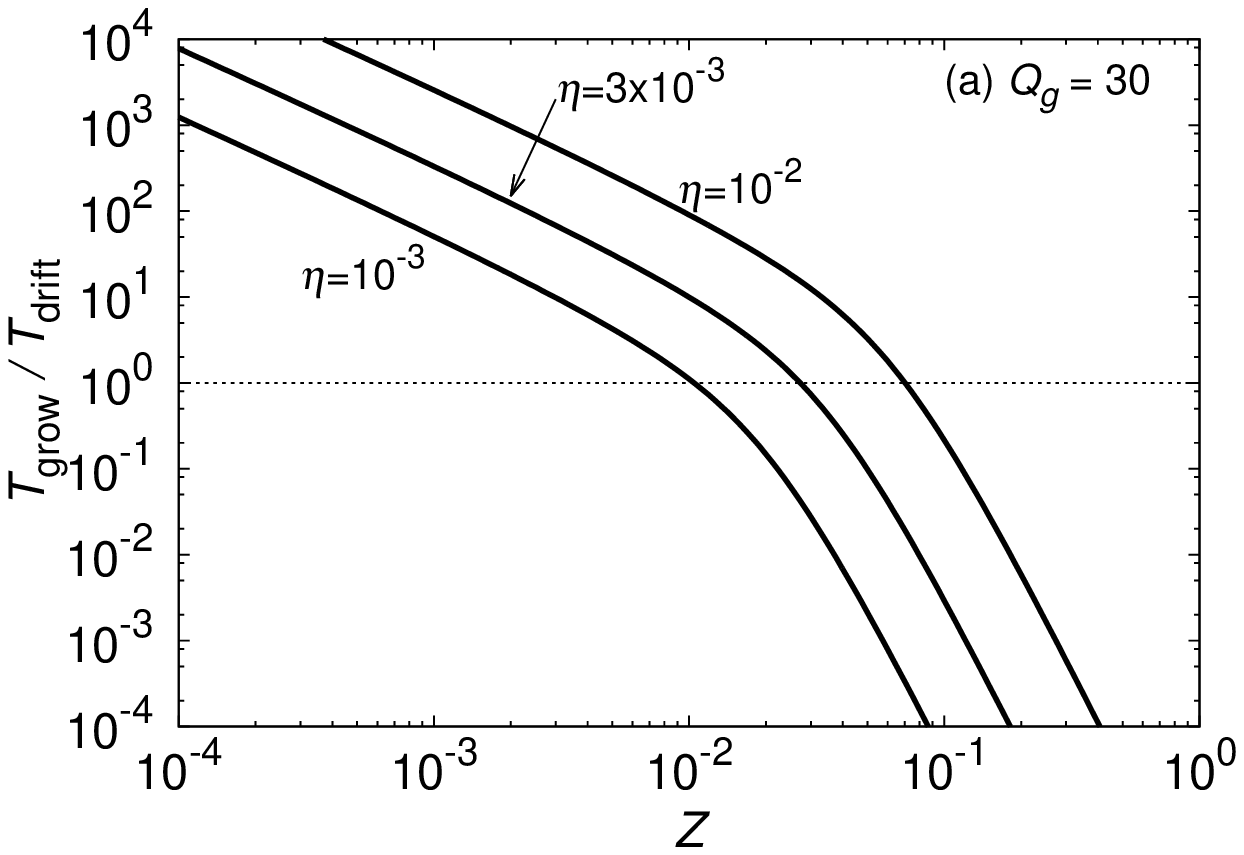} 
\plotone{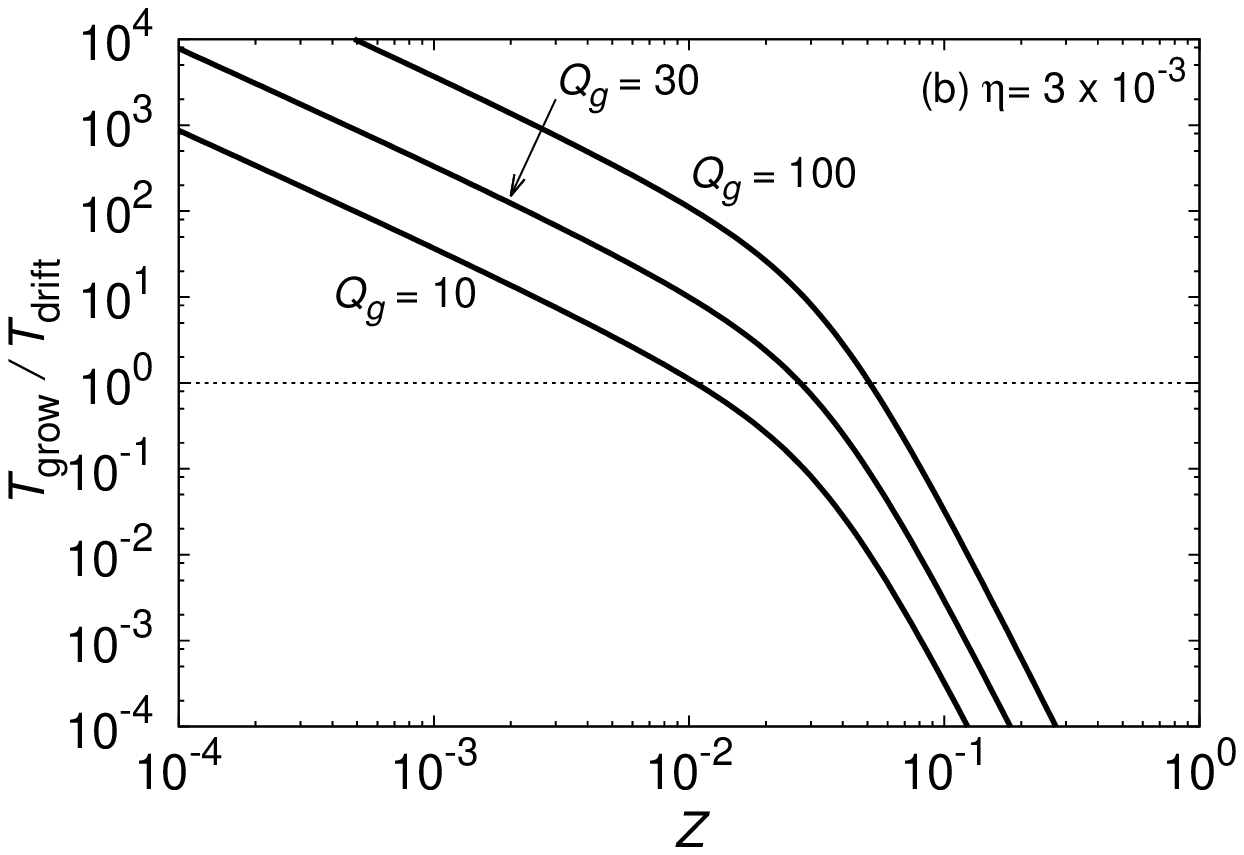} 
\caption{Ratio of the growth time to the radial drift time, $T_{\rm
    grow}/T_{\rm drift}$, against dust abundance $Z$. $(a)$ For models
  with a fixed Toomre parameter of the gas disk, $Q_{g}=30$. $(b)$ For
  models with a fixed deviation fraction of the gas velocity from the
  Keplerian velocity, $\eta=3 \times10^{-3}$. The horizontal dotted line
  indicates $T_{\rm grow}=T_{\rm drift}$.} 
\label{fig:Tg/Td}
\end{figure}

\subsection{Comparison of Timescales}

\label{sec:comptime}

\subsubsection{Minimum Dust Abundances}

Y11, SC11, and MKI12 show that dust layers are always unstable and subject to
SGI, but the growth rate for particles tightly coupled to the gas is strongly suppressed by
gas drag. The growth timescale can be larger than the other timescales, in
which the dust layer evolves significantly by other processes such as the
dispersal of the gas disk and the radial drift of the dust to the star. The
condition for SGI to be a relevant process for planetesimal formation
dictates that its growth timescale must be shorter than the other
evolution timescales. Y11 shows that the radial drift imposes the most
stringent condition for a wide range of disk parameters. We discuss the
condition for $T_{\rm grow}<T_{\rm drift}$. Because both $T_{\rm grow}$
and $T_{\rm drift}$ are inversely proportional to $T_{s}$, the ratio $T_{\rm grow}/T_{\rm drift}$ is independent of $T_{s}$; that is, the condition
$T_{\rm grow}<T_{\rm drift}$ is not affected by particle size. We
consider dependence of $T_{\rm grow}/T_{\rm drift}$ on the parameters $(\tilde{\eta},Z,Q_{g})$.

\begin{figure}
\epsscale{1.15} 
\plotone{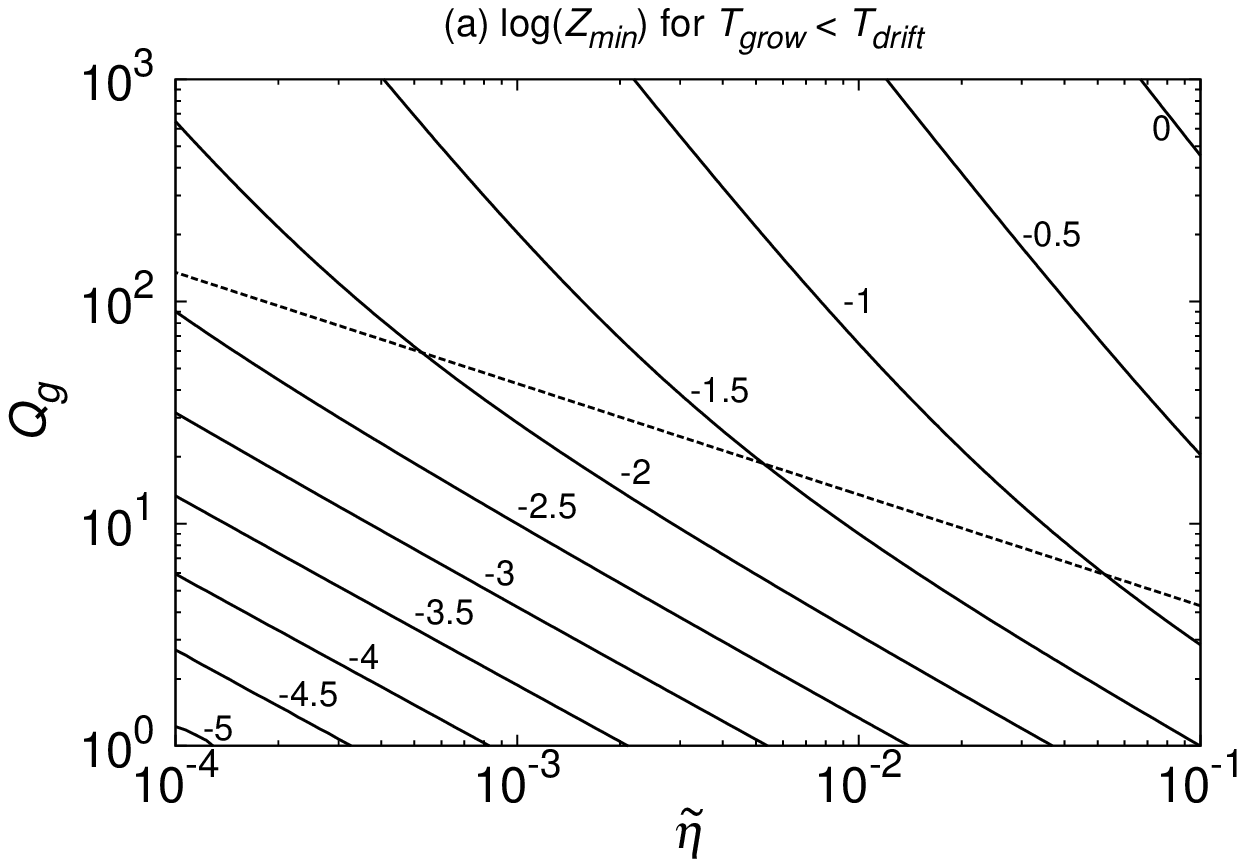} 
\plotone{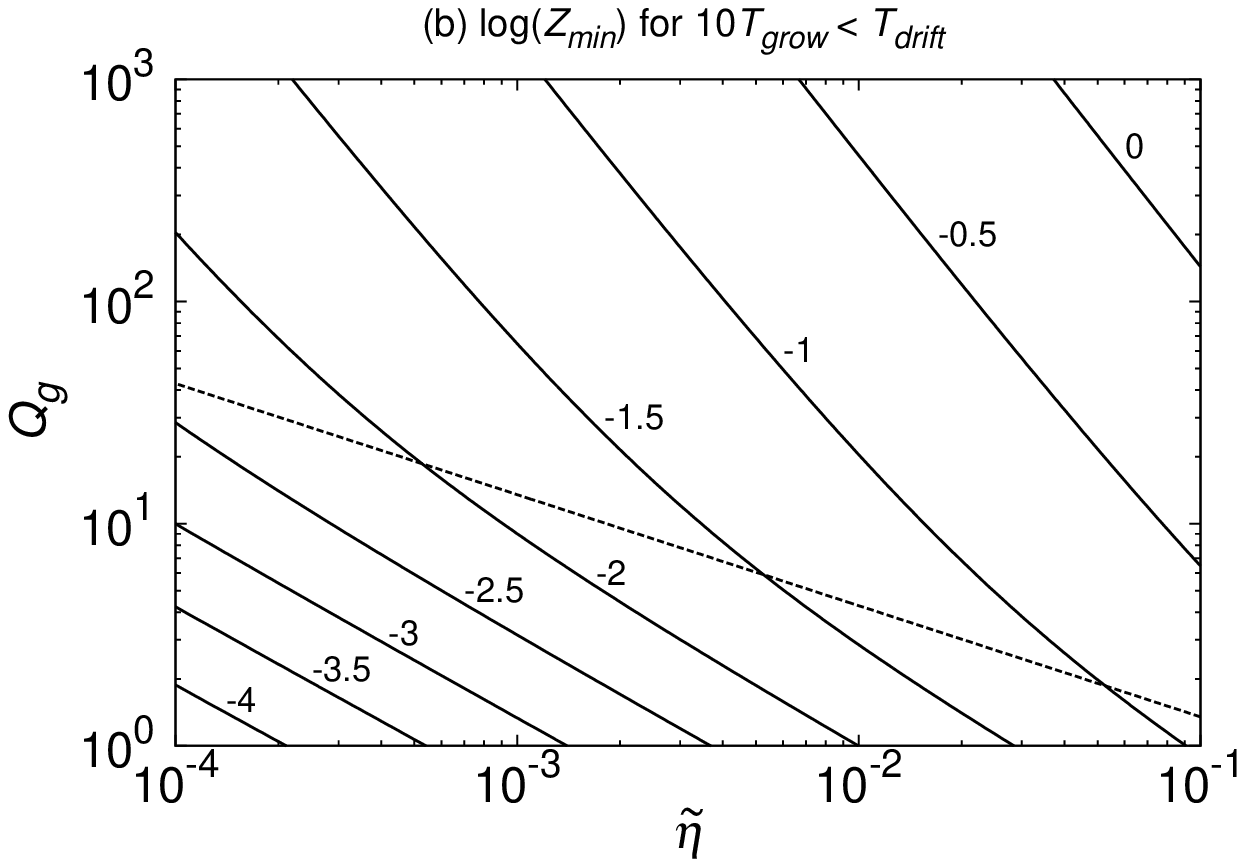}
\caption{Contour map of the minimum dust abundance $Z_{\rm min}$ on the
  $\tilde{\eta}$-$Q_{g}$ plane. 
$(a)$ $Z_{\rm min}$ required for the condition $T_{\rm grow} <
  T_{\rm drift}$. The contours are labeled by log$(Z_{\rm min})$. The
dashed line shows the locus of $Z_{\rm min} =(C_{\rm eff}\tilde{\eta})^{1/2}$
(i.e., $f_{\rm mid}\approx1$ when $Z=Z_{\rm min}$). $(b)$ $Z_{\rm
    min}$ required for the condition $10 T_{\rm grow} < T_{\rm drift}$. 
}
\label{fig:Zmin}
\end{figure}

First, we examine the limiting cases of $Z\ll(C_{\rm eff}\tilde{\eta})^{1/2}$ and
$Z\gg(C_{\rm eff}\tilde{\eta})^{1/2}$ \ ($f_{\rm mid}\ll1$ and $f_{\rm mid}\gg1$). In both limits,
\begin{eqnarray}
\lefteqn{ \frac{T_{\rm grow}}{T_{\rm drift}} } & & \nonumber \\
& \approx &
\left\{
\begin{array}
[c]{ccc}
2C_{\eta}^{-1}\left(  C_{1}C_{\rm eff}\right)  ^{2/3}\tilde{\eta}^{5/3}Q_{g}
^{2}Z^{-4/3} & \mathrm{for} & Z \ll (C_{\rm eff}\tilde{\eta})^{1/2} \\
2C_{\eta}^{-1}(C_{2} C_{\rm eff})^{3}\tilde{\eta}^{4}Q_{g}^{2}Z^{-6} &
\mathrm{for} & Z \gg (C_{\rm eff}\tilde{\eta})^{1/2}
\end{array}
\right.  \ . \nonumber \\
\label{eq:Tg/Td}
\end{eqnarray}
For $Z\ll(C_{\rm eff}\tilde{\eta})^{1/2}$, $T_{\rm grow}/T_{\rm drift}$
is proportional to $Z^{-4/3}$, while for $Z\gg(C_{\rm eff}\tilde{\eta})^{1/2}$
it rapidly decreases as $T_{\rm grow}/T_{\rm drift} \propto Z^{-6}$. In
Figure \ref{fig:Tg/Td}, $T_{\rm grow}/T_{\rm drift}$ is plotted against
$Z$ for various values of $\tilde{\eta}$ and $Q_{g}$. For $Z\gg(C_{\rm eff}
\tilde{\eta})^{1/2}\sim10^{-2}$, it is evident that $T_{\rm grow}/T_{\rm
  drift}$ rapidly decreases. For sufficiently large values of $Z$,
$T_{\rm grow}/T_{\rm drift}$ is less than unity, indicating that SGI
operates faster than the radial drift of the dust. From Figure \ref{fig:Tg/Td}, the minimum dust
abundance required for SGI, $Z_{\rm min}$, is determined as $Z$ satisfying
$T_{\rm grow}/T_{\rm drift}=1$. The minimum dust abundance $Z_{\rm min}$
ranges between $10^{-2}$ and $10^{-1}$ for parameters
$(\tilde{\eta},Q_{g})$ adopted in Figure \ref{fig:Tg/Td}.

In Figure \ref{fig:Zmin}(a), the minimum dust abundance $Z_{\rm min}$
required for $T_{\rm grow} < T_{\rm drift}$ is plotted as
contours on the $\tilde{\eta}$-$Q_{g}$ plane. The locus of
$Z_{\rm min}=(C_{\rm eff}\tilde{\eta})^{1/2}$ is represented as a dashed
line on the $\tilde{\eta}$-$Q_{g}$ plane. On this line, the midplane
dust-to-gas ratio $f_{\rm mid}$ becomes close to unity when $Z=Z_{\rm
  min}$. Substitution of $Z=(C_{\rm eff} \tilde{\eta})^{1/2}$ into
Equation (\ref{eq:Tg/Td}) shows that the condition 
$T_{\rm grow}/T_{\rm drift}=1$ (i.e., $Z=Z_{\rm min}$) becomes
$Q_{g}\approx\tilde{\eta}^{-1/2}$. Thus, the dashed line in Figure
\ref{fig:Zmin}(a) is linear with a slope of $-1/2$. Above this line, $Z_{\min
}>(C_{\rm eff}\tilde{\eta})^{1/2}$ (and $f_{\rm mid}>1$ for $Z=Z_{\rm min}$); below
it, $Z_{\rm min}<(C_{\rm eff}\tilde{\eta})^{1/2}$ (and $f_{\rm mid}<1$). The
dependence of $Z_{\rm min}$ on $(\tilde{\eta},Q_{g})$ is expressed as
\begin{equation}
Z_{\rm min}\approx\left\{
\begin{array}
[c]{ccc}
(8C_{\eta}^{-3}C_{1}^{2}C_{\rm eff}^{2}\tilde{\eta}^{5}Q_{g}^{6})^{1/4} &
\mathrm{for} & Q_{g}\ll\tilde{\eta}^{-1/2}\\
(2C_{\eta}^{-1}C_{2}^{3}C_{\rm eff}^{3}\tilde{\eta}^{4}Q_{g}^{2})^{1/6} &
\mathrm{for} & Q_{g}\gg\tilde{\eta}^{-1/2}
\end{array}
\right.  ~.
\label{eq:zmin}
\end{equation}
Larger dust abundances are required for larger $\tilde{\eta}$ (i.e., hotter
gas disks) and larger $Q_{g}$ (i.e., less massive gas disks). In standard
models for protoplanetary disks, $\tilde{\eta}=10^{-3}-10^{-2}$, and
$Q_{g}=10-1000$ (see Figure \ref{fig:model-qg-eta} below). Thus, if the dust
abundance $Z$ is larger than $0.1$, such dusty disks operate SGI. If the gas
disk is so cold that $\tilde{\eta}$ is as small as $10^{-3}$ and if it is so
massive that $Q_{g}$ is as small as $10$, then even a standard value $Z=0.01$
is sufficiently large to operate SGI.

\begin{figure}
\epsscale{1.15} 
\plotone{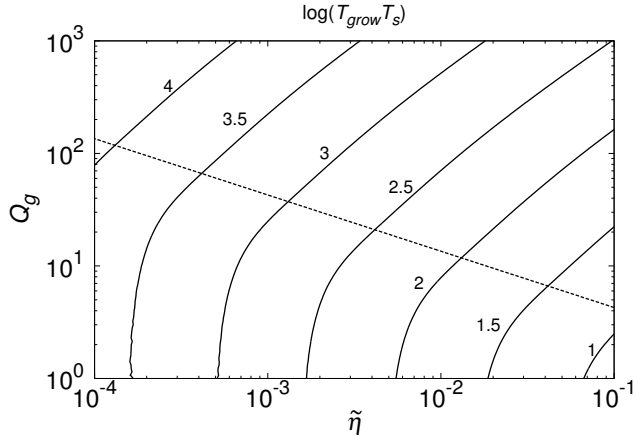}
\caption{Contour map of the growth time multiplied by the stopping time,
  $T_{\rm grow}T_{s}$, on the $\tilde{\eta}$-$Q_{g}$ plane. The contours
  are labeled by log$(T_{\rm grow}T_{s})$. In calculating $T_{\rm
    grow}(\tilde{\eta},Q_{g})$, the minimum dust abundances $Z_{\rm
    min}(\tilde{\eta},Q_{g})$ are used. The dashed line shows the locus 
  of $Z_{\rm min} = (C_{\rm eff} \tilde{\eta})^{1/2}$. } 
\label{fig:Tgrow}
\end{figure}

\subsubsection{Dependence of $Z_{\rm min}$ on Some Model Parameters}

The growth time $T_{\rm grow}$ is just a time for density
  perturbations to increase $e$-fold. Because many $e$-folds are needed
  for a significant density increase, actual condition for SGI would be
  $R T_{\rm grow} < T_{\rm drift}$, where $R>1$ is the required number
  of $e$-folding. From Equation (\ref{eq:Tg/Td}) the minimum dust abundance $Z_{\rm min}$ scales as
  $Z_{\rm min} \propto R^{3/4}$ for $Q_g \ll (R \eta)^{-1/2}$ and
  $Z_{\rm min} \propto R^{1/6}$ for $Q_g \gg (R \eta)^{-1/2}$. Figure
  \ref{fig:Zmin}(b) shows $Z_{\rm min}$ derived for the condition $10 T_{\rm
    grow} < T_{\rm drift}$. For less massive gas disks ($Q_g \gg (R
  \eta)^{-1/2}$; above the dashed line in Figure \ref{fig:Zmin}), $Z_{\rm
    min}$ depends only weakly on $R$. For massive gas disks ($Q_g \ll (R
  \eta)^{-1/2}$), required $Z_{\rm min}$ is higher for larger $R$
  ($Z_{\rm min} \propto R^{3/4}$), but would not be too high (i.e.,
  $Z_{\rm min} \la 0.1$ even for $R=10$; see the region below the dashed line in
  Figure \ref{fig:Zmin}(b)). In the following discussions, we use the condition
  $T_{\rm grow} < T_{\rm drift}$. 

In this paper, it is assumed that about 20\% of the dust
accretion energy is used for turbulence excitation ($C_{\rm
  eff}=0.19$). This value is determined by comparison with the numerical
simulations of turbulence excited via KH instability by Johansen et
al. (2006). Because their simulation was two-dimensional, the realistic
value of $C_{\rm eff}$ could be different. For example,
three-dimensional simulations by Lee et al. (2010) show that the
critical Richardson number for KH instability increases with $Z$,
implying that $C_{\rm eff}$ also increases with $Z$ (see discussion in
Section 5.2.1 of T12). As shown in Equation (\ref{eq:zmin}), $Z_{\rm min} \propto
C_{\rm eff}^{1/2}$. For the maximum efficiency ($C_{\rm eff}=1$),
$Z_{\rm min}$ would be about 2 times larger than our estimate. 

\begin{figure}
\epsscale{1.15} 
\plotone{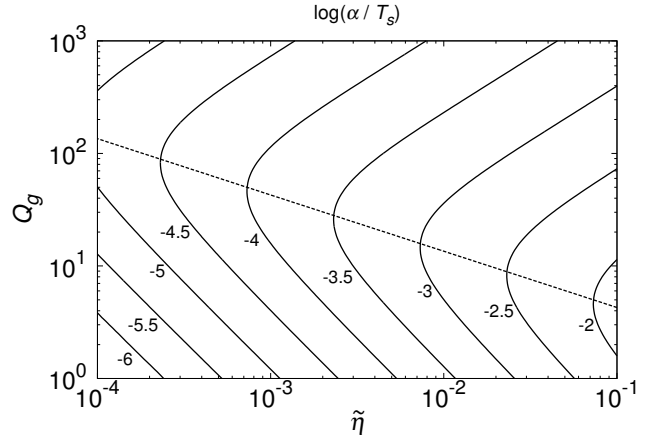}
\caption{Contour map of the diffusion parameter divided by the stopping
  time, $\alpha/T_{s}$, on the $\tilde{\eta}$-$Q_{g}$ plane for the dust
  layer with $Z_{\rm min}$. The contours are labeled by
  log$(\alpha/T_{s})$. The dashed line shows the locus of $Z_{\rm min} =
  (C_{\rm eff} \tilde{\eta})^{1/2}$. } 
\label{fig:alpha}
\end{figure}

\subsection{Conditions on $T_{\rm grow}$, $\alpha$, and $\lambda$}

In the previous subsection, the condition for $T_{\rm grow}
<T_{\rm drift}$ was discussed. Next, we discuss other constraints
required for SGI operation. First, we consider the condition for the growth time to be less than the
disk life time. The growth time multiplied by the stopping time,
$T_{\rm grow}T_{s}$, is plotted on the $\tilde{\eta}$-$Q_{g}$ plane
in Figure \ref{fig:Tgrow}. The values of $T_{\rm grow}(\tilde{\eta
},Q_{g})$ are calculated using the minimum dust abundance $Z=Z_{\rm min}
(\tilde{\eta},Q_{g})$, the values of which differ according to $(\tilde{\eta
},Q_{g})$ as shown in Figure \ref{fig:Zmin}. For larger values of $Z$,
$T_{\rm grow}$ is shorter ($T_{\rm grow}\propto
Z^{-4/3}$ for $Z<(C_{\rm eff}\tilde{\eta})^{1/2}$ and $T_{\rm grow
}\propto Z^{-4}$ for $Z>(C_{\rm eff}\tilde{\eta})^{1/2}$). Because
$T_{\rm grow}$ is inversely proportional to $T_{s}$, the contours
are labeled by $\log(T_{\rm grow}T_{s})$. Thus, the growth
timescale is obtained by dividing the value in Figure \ref{fig:Tgrow} by
$T_{s}$. For $\tilde{\eta}>10^{-3}$ and $Q_{g}<10^{2}$,
$T_{\rm grow}$ is less than $10^{3}T_{s}^{-1}\Omega_{\rm K}^{-1}$. Thus
at 1AU, the dust layer composed of $T_{s}\ga 10^{-3}$ particles (size
$a\ga 1$mm) with the dust abundance $Z_{\rm min}$ operates SGI within a disk life
time of $\sim1$Myr.

Figure \ref{fig:alpha} shows the diffusion parameter $\alpha$ due to
turbulence induced in the dust layer with the abundance $Z_{\rm min}$. Because
$\alpha$ is proportional to $T_{s}$, the contours are labeled by $\log
(\alpha/T_{s})$. Thus, the $\alpha$ value is obtained by multiplying the
value in Figure \ref{fig:alpha} by $T_{s}$. For $\tilde{\eta}>10^{-3}$ and
$Q_{g}<10^{2}$, $\alpha$ is larger than $10^{-4}T_{s}$, except for very small
$Q_{g}\la10$ and $\tilde{\eta}\la3\times10^{-3}$. If the disk is turbulent due
to other mechanisms than that originating from the dust layer and if turbulent
diffusion is stronger than that shown in Figure \ref{fig:alpha}, then
the required value for $Z$ would be higher. Even in the dead zone where
MRI is not active, the gas could have turbulent motion, causing
diffusion of the dust (Fleming \&\ Stone 2003). Okuzumi \& Hirose (2011)
show that lower values of the diffusion coefficient in the dead zone are
realized for a wider dead zone or weaker vertical magnetic
fields. For example, if the plasma $\beta$ (the ratio of gas
pressure to magnetic pressure) is greater than $3 \times 10^6$, the
diffusion coefficient due to MRI turbulence is less than $10^{-5}$ (see
model X1b of Okuzumi \& Hirose 2011). In such a weakly magnetized disk
and for $T_s >0.1$ particles, turbulence originating in the dust layer is
stronger than MRI turbulence in the dead zone.

\begin{figure}
\epsscale{1.15} 
\plotone{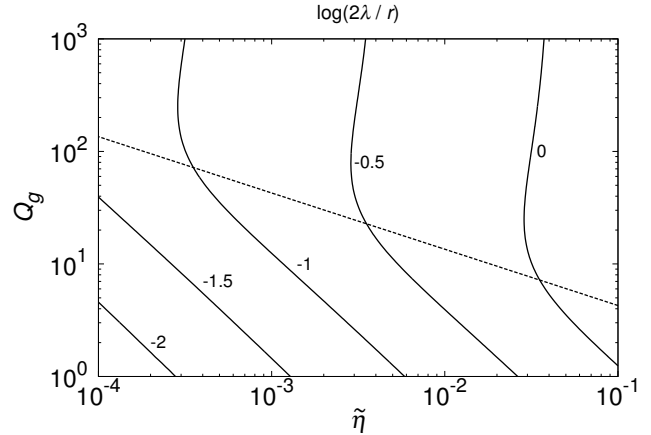}
\caption{Contour map of the normalized wavelength of the most unstable
  mode $2 \lambda/ r$ on the $\tilde{\eta}$-$Q_{g}$ plane for the dust
  layer with $Z_{\rm min}$. The contours are labeled by log$(2 \lambda/
  r)$. The dashed line shows the locus of $Z_{\rm min} = (C_{\rm eff} 
\tilde{\eta})^{1/2}$.}
\label{fig:lambda}
\end{figure}

The wavelength of the most unstable mode must be smaller than the disk radius.
The wavelength $\lambda$ is given in Equation (56) of Y11. The ratio of
$\lambda$ to half of the disk radius $r$ is
\begin{equation}
\frac{2\lambda}{r}\approx\frac{4\pi\alpha Q_{g}\tilde{\eta}^{1/2}}{C_{\eta} Z T_{s}}~,
\end{equation}
and is plotted in Figure \ref{fig:lambda} for disks with the minimum dust
abundances $Z_{\rm min}$. Note that $2\lambda/r$ does not depend on $T_{s}$,
because $\alpha$ is proportional to $T_{s}$. Figure \ref{fig:lambda} shows
that $\lambda$ is smaller than $r/2$ if $\tilde{\eta}$ is less than
$3\times10^{-2}$, although it is still comparable to $r/2$ even for $\tilde
{\eta}$ as small as $3\times10^{-3}$. In the analysis by Y11, SC11,
and MKI12, the gas disk is assumed to behave as a stationary
background, and its velocity profiles 
are not affected by the gravity of the dust layer. However, the response of
the gas to perturbations longer than the gas scale height is not clear. For
perturbations with smaller wavelengths to be unstable, larger values of $Z$
than those shown in Figure \ref{fig:Zmin} are required. The wavelength varies as
$\lambda\propto Z^{-1/3}$ for $Z<(C_{\rm eff}\tilde{\eta})^{1/2}$ and
$\lambda\propto Z^{-3}$ for $Z>(C_{\rm eff}\tilde{\eta})^{1/2}$. In the latter
regime ($Z>(C_{\rm eff}\tilde{\eta})^{1/2}$), an increase by a factor of
2 in $Z$ results in an order of magnitude decrease in $\lambda$, while in the former
regime, reduction in the wavelength requires a large increase in $Z$.

\begin{figure}
\epsscale{1.15} 
\plotone{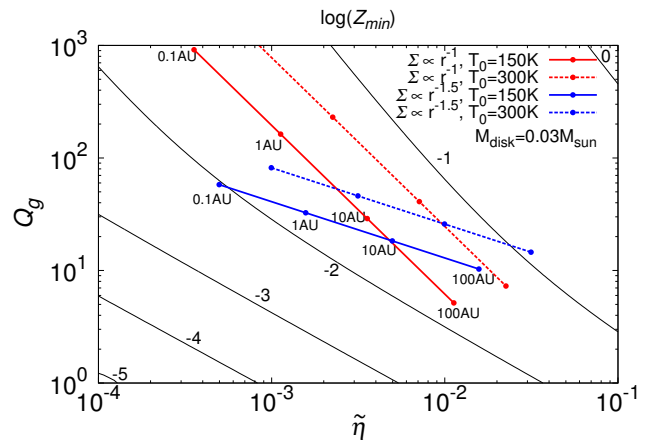}
\caption{Models of the gas disks are plotted on the
  $\tilde{\eta}$-$Q_{g}$ plane. Red and blue lines represent models with
  shallow and steep density profiles, $\Sigma\propto r^{-1}$ and
  $\Sigma\propto r^{-1.5}$, respectively. Solid and dashed lines
  indicate models of cold and hot gas disks, $T_{0}=150$K and
  $T_{0}=300$K, respectively. Dots on the lines show locations on the
  disks, $0.1$, 1, 10, and 100AU. The disk mass is $M_{\rm disk}=0.03
  M_{\sun}$. Contours show $\log(Z_{\rm min})$.} 
\label{fig:model-qg-eta}
\end{figure}

\section{Minimum Dust Abundances for Various Disk Models}

\label{sec:diskmodel}

In the previous section, minimum dust abundances for SGI were obtained for
given parameters $(\tilde{\eta},Q_{g})$. In this section, we consider
several disk models. We adopt power-law profiles for density and temperature
distributions of the gas disk, such that
\begin{equation}
\Sigma_{g}=\Sigma_{g,0}r_{\rm AU}^{-p}~,
\end{equation}
\begin{equation}
T=T_{0}r_{\rm AU}^{-q}~,
\end{equation}
where $r_{\rm AU}$ is the distance from the star measured in AU. In such a disk,
$Q_{g}\propto r_{\rm AU}^{(2p-q-3)/2}$ and $\tilde{\eta}\propto r_{\rm AU}^{1-q}$. We
consider two cases of hot and cold gas disks, $T_{0}=300$K and $150$K, while
we fix the power-law index $q=1/2$. For the density profile, $p=1.0$ and $1.5$
are adopted and the mass of the gas disk inside 100 AU is varied within
$M_{\rm disk}=10^{-2}-10^{-1}M_{\sun}$. Figure \ref{fig:model-qg-eta}
shows the variability in the values of $\tilde{\eta}$ and $Q_{g}$
between the models and with $r$. In this figure, models with $M_{\rm
  disk}=3\times10^{-2}M_{\sun}$ are shown. For various values of $M_{\rm disk}$, $Q_{g}$ scales as $Q_{g}\propto M_{\rm disk}$.

\begin{figure}
\epsscale{1.15} 
\plotone{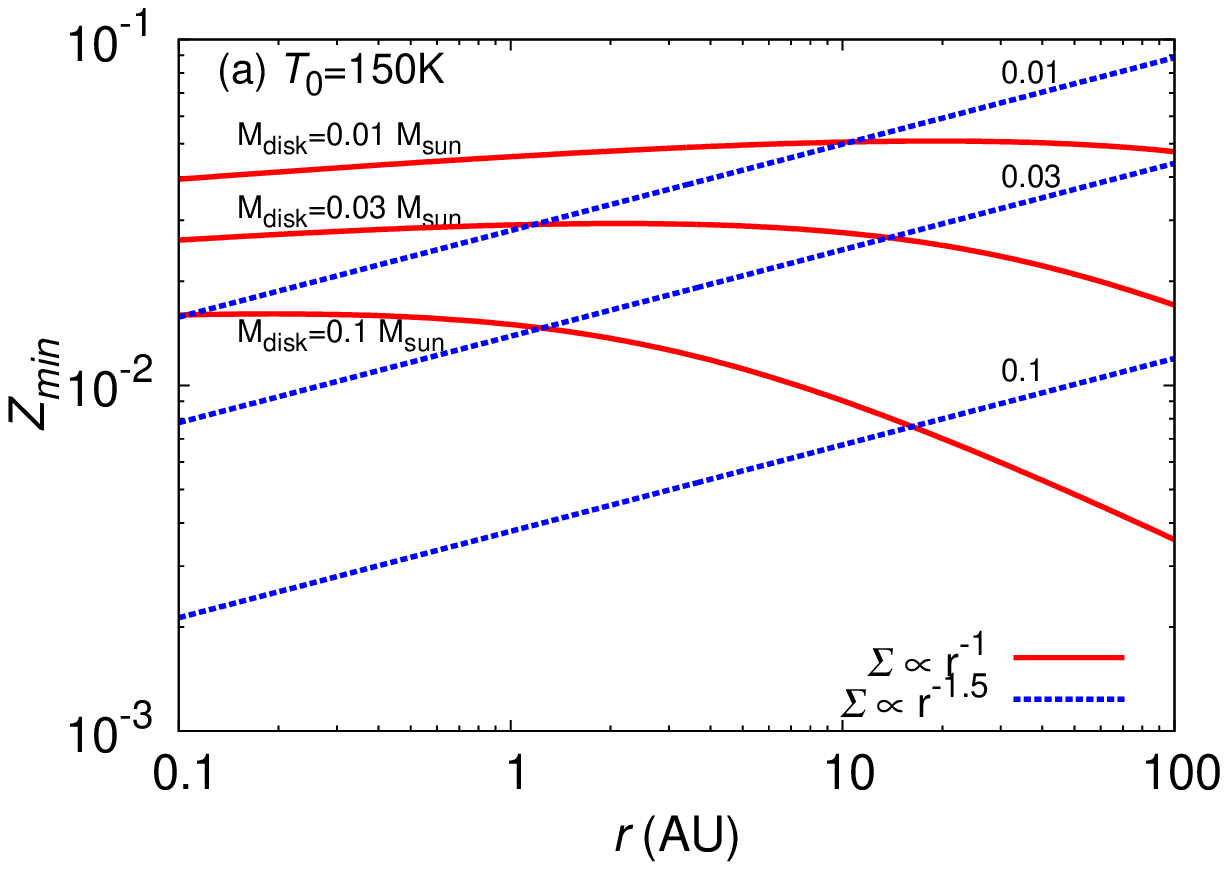} 
\plotone{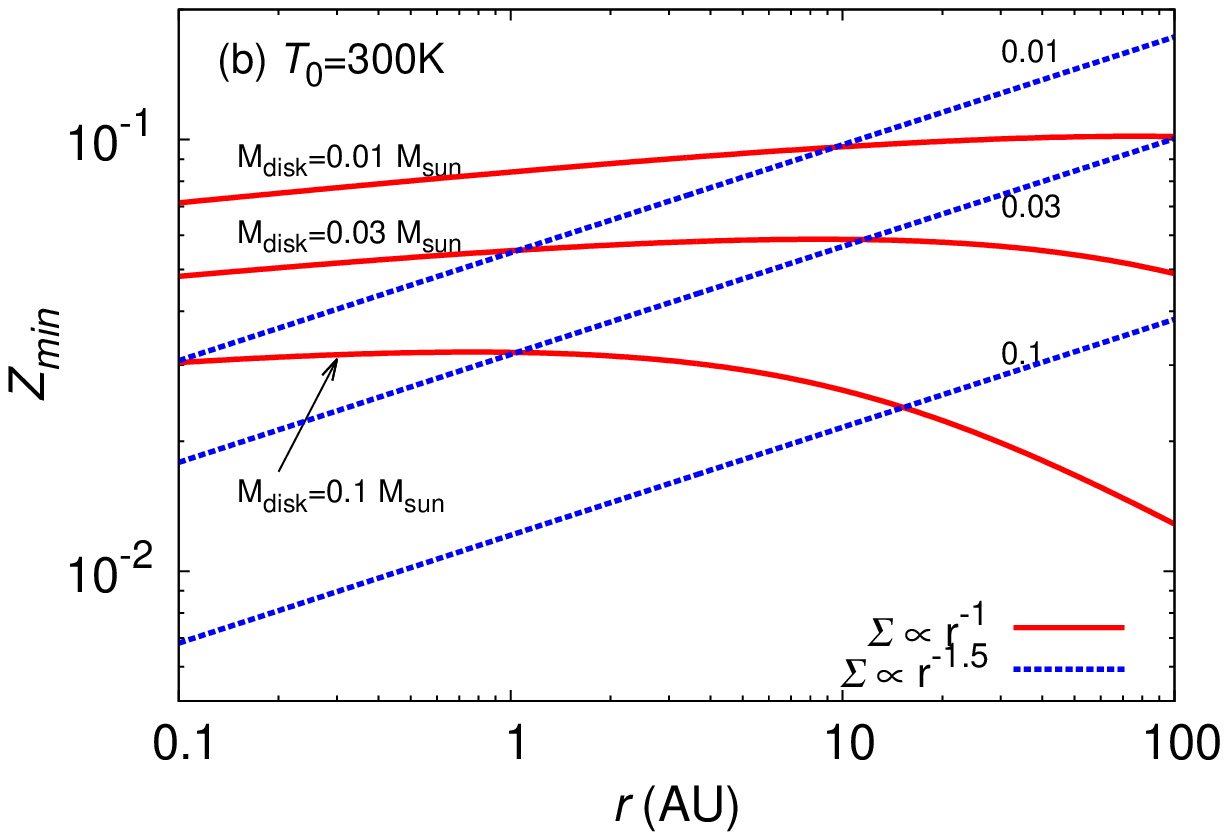} 
\caption{Minimum dust abundances $Z_{\rm min}$ for various disk
  models. $(a)$ Cold disks ($T_{0}=150$K). $(b)$ Hot disks
  ($T_{0}=300$K). Red solid and blue dashed lines represent models with
  shallow and steep density profiles, $\Sigma_{g} \propto r^{-1}$ and
  $\Sigma_{g} \propto r^{-1.5}$, respectively.} 
\label{fig:diskmodel}
\end{figure}

Figure \ref{fig:diskmodel} shows the required minimum dust abundances $Z_{\min
}$ against $r_{\rm AU}$ for various disk models. For the steep density profile
($p=1.5$; blue dashed lines), $Z_{\rm min}\propto r_{\rm AU}^{1/4}$. (In the regime of $Q_{g}\ll
\tilde{\eta}^{-1/2}$ of Equation (\ref{eq:zmin}), $Z_{\rm min}\propto\tilde{\eta
}^{5/4}Q_{g}^{3/2}\propto r_{\rm AU}^{1/4}$; in the $Q_{g}\gg\tilde{\eta
}^{-1/2}$ regime, $Z_{\rm min}\propto\tilde{\eta}^{2/3}Q_{g}^{1/3}\propto
r_{\rm AU}^{1/4}$). SGI is more viable at the inner part of the
disk. For $p=1.0$ (red solid lines), $Z_{\rm min}\propto r_{\rm
  AU}^{1/12}$ only weakly depends on $r$ (in the $Q_{g} \gg
\tilde{\eta}^{-1/2}$ regime). The outer part of the disk with 
$M_{\rm disk}=10^{-1}M_{\sun}$ and $p=-1$ is in the $Q_{g}\ll\tilde{\eta}^{-1/2}$
regime, and $Z_{\rm min}$ decreases as $Z_{\rm min} \propto
r^{-1/2}$. Even in the hot gas disks ($T_{0}=300$K) with a mass as small
as $M_{\rm disk}=10^{-2}M_{\sun}$, the required dust abundances for SGI
are less than $0.1$. If the temperature at the disk midplane is as cold
as $T_{0}\approx150$K, as expected in passive disks that are heated only
by stellar radiation (Chiang \& Goldreich 1997; Tanaka et al. 2005), SGI
operates for $Z_{\min}<0.05$ at 10AU of disks with $M_{\rm
  disk}=10^{-2}M_{\sun}$.

\section{Discussion}

Analysis in this paper assumes that turbulence is induced in the dust
layer. If the dust abundance at the midplane $f_{\rm mid}$ is larger
than unity ($Z\ga(C_{\rm eff} \tilde{\eta})^{1/2}$), turbulence weakens
with increasing $Z$ as 
$\alpha\propto Z^{-2}$ (Equation (\ref{eq:alpha})). The growth timescale of SGI
rapidly decreases as $T_{\rm grow}\propto Z^{-4}$ (Equation (\ref{eq:tgrow2})). A slight
increase in $Z$ makes SGI viable. Y11 also estimated the minimum dust
abundance $Z_{\eta}$ required for SGI (Equation (69) of Y11), assuming that the
turbulent diffusion parameter of a dense dust layer ($f_{\rm mid}\ga1$) was
$\alpha_{\eta}\approx T_{s}\tilde{\eta}$ (Equation (68) of Y11), independent
of $Z$. This estimate for $\alpha_{\eta}$ adopted by Y11 corresponds to
the maximum value of $\alpha(Z)$ at $Z\approx(C_{\rm eff}
\tilde{\eta})^{1/2}$ in our estimate (Equation (\ref{eq:alpha})). If
$\alpha$ had a constant value $\alpha_{\eta}$ for all $Z$, the condition
$T_{\rm grow}<T_{\rm drift}$ would be
\begin{equation}
Z>Z_{\eta}\sim \eta^{5/6} Q_{g}^{2/3} ~,
 \label{eq:Zeta}
\end{equation}
where we used $f_{\rm mid}=Z h_{g}/h_{d}$ and $h_{d}/h_{g}\sim\eta^{1/2}$. Note
that this Equation (\ref{eq:Zeta}) differs from Equation (69) of Y11,
because we use Equation (\ref{eq:Tdrift}), in which $T_{\rm drift}\propto f_{\rm mid}$, while Y11 
assumes $T_{\rm drift}\propto f_{\rm mid}^{2}$ (see discussion following Equation
(\ref{eq:vdr}) below). As discussed in T12, accumulation of the dust increases
the inertia of the dust layer and decelerates the inward drift of the dust,
resulting in weaker turbulence. Considering $\alpha \propto
Z^{-2}$, the condition $T_{\rm grow}<T_{\rm drift}$ becomes
\begin{equation}
Z>Z_{\rm min}\sim \eta^{2/3} Q_{g}^{1/3} ~,
\label{eq:Zmin2}
\end{equation}
as shown in Equation (\ref{eq:zmin}). We ignore numerical coefficients such as
$C_{\rm eff}$ and $C_{\eta}$. These two conditions (Equations (\ref{eq:Zeta}) and
(\ref{eq:Zmin2})) coincide when $f_{\rm mid}=1$ (i.e., $Z_{\eta}=Z_{\rm min}\approx
\eta^{1/2}$) or when $Q_{g}=Q_{g,0}(\eta)\equiv\eta^{-1/2}$ (i.e., on the
dashed line in Figures \ref{fig:Zmin}-\ref{fig:lambda}). If the disk's
$Q_{g}$ is greater than $Q_{g,0}$, then the latter condition (Equation (\ref{eq:Zmin2}))
permits smaller dust abundances for SGI by a factor of $Z_{\rm min}/Z_{\eta}
=(Q_{g}/Q_{g,0})^{-1/3}$. For example, at 1AU on a disk with $M_{\rm disk}
=0.03M_{\sun}$, $\Sigma_{g}\propto r^{-1},$ and $T_{0}=300$K, Equation
(\ref{eq:Zmin2}) gives $Z_{\rm min}=0.11$, which is smaller by a factor of 2 than
the value $Z_{\eta}=0.23$ predicted by Equation (\ref{eq:Zeta}). Considering
the $Z$ dependence of $\alpha$ reduces the required minimum value
$Z_{\rm min}$ for SGI as compared with that derived by Y11.

\section{Summary}

We analyze the condition for SGI occurrence in the dust layer. The growth timescale
of the instability must be shorter than the radial drift timescale of the
dust. The growth timescale decreases as turbulence weakens. The necessary
condition for SGI is obtained by considering turbulence induced in the dust
layer. Using the turbulence strength estimated from the energy supply rate
from the accreting dust, the minimum dust abundances for SGI, $Z_{\rm min}$, are
derived as a function of the Toomre $Q_{g}$ parameter of the gas
  disk and the deviation fraction $\eta$ of the gas velocity from the
Keplerian value. If the dust particles are small and their stopping time
is less than the Keplerian time 
($T_{s}<1$), $Z_{\rm min}$ is independent of $T_{s}$. For disks with $Q_{g}
\sim10^{2}$ and $\eta\sim10^{-2}$, SGI occurs if $Z\ga0.1$. The required $Z$
decreases with decreasing $Q_{g}$ and $\eta$, becoming as small as
$0.01$ for $Q_{g}\sim10$ and $\eta\sim10^{-3}$. Such an increase in $Z$
from the solar abundance is expected to occur through several processes,
including the radial drift of the dust and dispersal of the gas from the disk
(Youdin \& Shu 2002; Takeuchi \& Lin 2002; Takeuchi et
al. 2005). Therefore, SGI provides a possible route for small particles
to directly collapse to planetesimals in an initially laminar 
disk. If the gas disk were globally turbulent via MRI, for example, the
required $Z$ would be larger than $Z_{\rm min}$ estimated in this paper.

\acknowledgements
We thank Takayuki Muto, Satoshi Okuzumi, Hidekazu Tanaka, Shugo
Michikoshi, and Naoki Ishitsu for useful discussions. We also thank
  an anonymous referee for helpful comments. This work was
supported by Grants-in-Aid for Scientific Research, Nos. 20244013 and
20540232 from the Ministry of Education, Culture, Sports, Science, and
Technology, Japan.

\appendix

\section{Approximate Expressions of Diffusion Parameter and Radial
Drift Rate}
\label{sec:app-approx}

In this Appendix, we evaluate an approximate expression of the turbulent
diffusion parameter $\alpha$ for small particle limit ($T_{s}\ll1$) according
to T12. We consider turbulence excited via KH or streaming
  instabilities of the dust layer. The strength of the turbulence is
controlled by the dust accretion rate. The dust particles accrete toward
the star either due to the gas drag on individual particles (Nakagawa et
al. 1986) or collective drag exerted on the entire dust layer
(Weidenschilling 2003; Youdin \& Chiang 2004). Individual drag dominates
if the dust-to-gas ratio in the dust layer is less than unity ($f_{\rm
  mid}\la1$), while collective drag is important if $f_{\rm mid}\ga1$.

First, we consider individual drag (for $f_{\rm mid}\la1$). Using the formula
derived by Nakagawa et al. (1986; or Equation (9) of T12, hereafter
Equation (T9)), the accretion velocity of small dust particles
($T_{s}\la1$) is
\begin{equation}
v_{d,r}=-2T_{s}\eta v_{\rm K}~.\label{eq:vdr-ind}
\end{equation}
To satisfy angular momentum conservation, the gas moves outward with the
velocity $v_{g,r}=-v_{d,r} \rho_d / \rho_g$, where $\rho_d$ ($\rho_g$) is
the dust (gas) density. The effective gravities (including the pressure
gradient force) acting on the dust and gas are
$g_{d}=-r\Omega_{\mathrm{K}}^{2}$ and $g_{g} = -(1-2\eta) r
\Omega_{\mathrm{K}}^{2}$, respectively. The net work done by these
effective gravities on a unit surface area of the dust layer is
calculated as $\Delta E_{\rm drag} \sim \int (\rho_{d} g_{d} v_{d,r} +
\rho_{g} g_{g} v_{g,r}) dz \sim \eta^{2} v_{\rm K}^{2} \Omega_{\rm K}
T_{s} \Sigma_{d}$ (Equation (T12)). The energy dissipation due to turbulence is estimated as
$\Delta E_{\rm turb}\sim\Sigma_{\rm layer}u_{\rm eddy}^{2}/\tau_{\rm eddy}$, where
$\Sigma_{\rm layer}$ is the surface density of the dust layer, and $u_{\rm eddy}$ and
$\tau_{\rm eddy}$ are the typical values for the velocity and turnover
time of the largest 
eddies, respectively. Adopting the so-called \textquotedblleft$\alpha$
prescription\textquotedblright, we evaluate $u_{\rm eddy}\sim\sqrt{\alpha}c_{g}$
and $\tau_{\rm eddy} \sim \Omega_{\rm K}^{-1}$ (Cuzzi et al. 2001). For
$f_{\rm mid}\la1$, the column density of the dust layer is dominated by the gas, and thus
$\Sigma_{\rm layer}\sim\Sigma_{g}h_{d}/h_{g}$. Here, $h_{d}/h_{g}$ is the ratio of
the scale heights of the dust layer and gas disk, and is given by
$h_{d}/h_{g}\approx\sqrt{\alpha/T_{s}}$ (Equation (T4)). Then, the energy
dissipation rate is $\Delta E_{\rm turb}\sim\sqrt{\alpha^{3}/T_{s}}h_{g}^{2}
\Omega_{\rm K}^{3}\Sigma_{g}$. Equating the energy supply and dissipation rates of
turbulence, with an efficiency parameter for the energy supply, $\Delta
E_{\rm turb}\sim C_{\rm eff}\Delta E_{\rm drag}$, gives
\begin{equation}
\alpha\sim(C_{\rm eff}\tilde{\eta}Z)^{2/3}T_{s}~.\label{eq:aplha-apx1}
\end{equation}

\begin{figure}
\epsscale{1.15} 
\plottwo{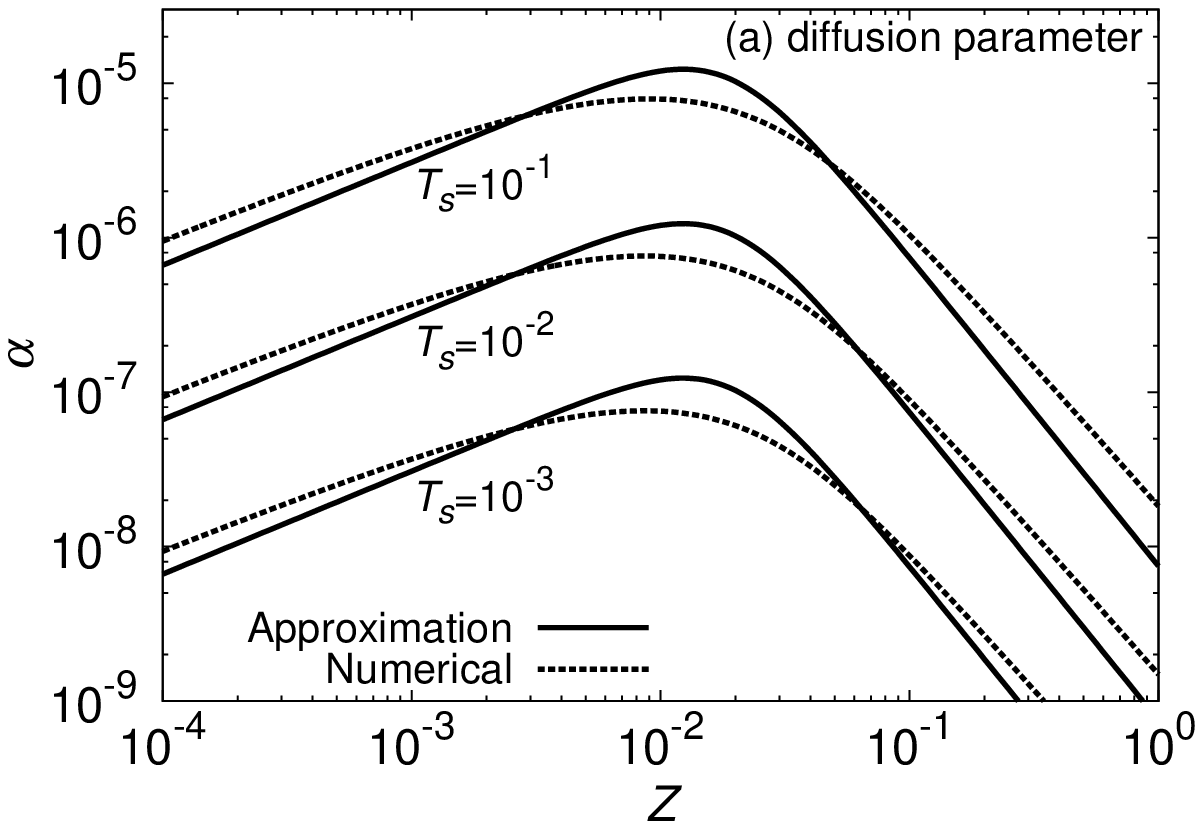}{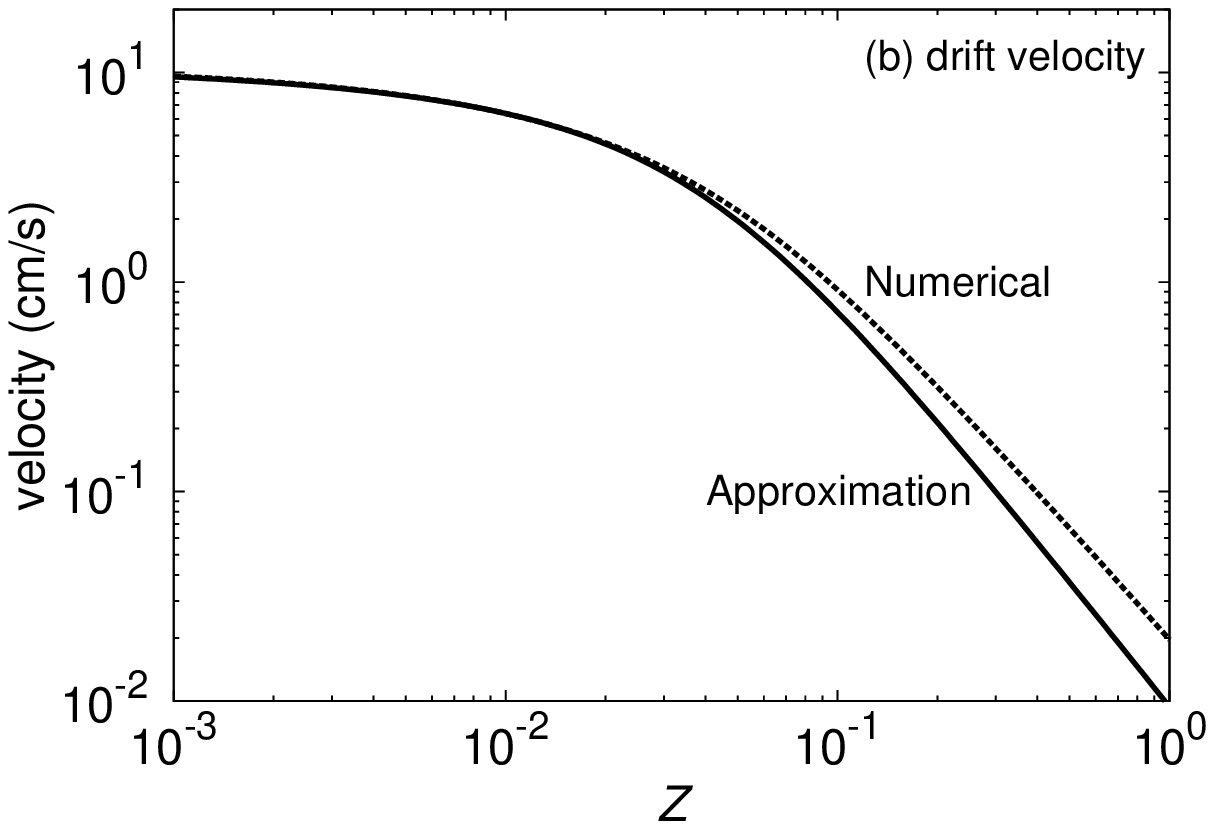}
\caption{Comparison of approximate expressions with numerical results
  for $(a)$ turbulent diffusion parameter $\alpha$ and $(b)$ dust radial
  drift velocity $v_{d,r}$. Solid lines represent approximate
  expressions (\ref{eq:alpha-apx3}) and (\ref{eq:vdr}), and dashed lines
  are numerically calculated by T12. $(a)$ Diffusion parameter $\alpha$
  is plotted for $T_{s}=10^{-3}$,$10^{-2}$, and $10^{-1}$. $(b)$ Radial
  drift velocity $v_{d,r}$ is plotted for $T_{s}=10^{-3}$. }
\label{fig:alp-app}
\end{figure}

The collective drag force in the $\theta$-direction on a unit surface area is
estimated by the plate drag approximation for $f_{\rm mid}\ga1$ as $P_{\theta
z}\sim\rho_{g}\nu\partial v_{g,\theta}/\partial z\sim-\rho_{g}\nu\eta
v_{\rm K}/h_{d}$, where $\nu$ is the turbulent viscosity, and the
velocity shear in the vertical direction $\partial v_{g,\theta}/\partial
z$ is estimated as $\eta v_{\rm K}/h_{d}$. This drag causes 
accretion of the dust layer and an outward motion of the upper gas layer. 
The dust layer loses energy $\Omega_{\rm K} r P_{\theta z}$ and the
upper gas layer gains energy $(1-\eta) \Omega_{\rm K} r P_{\theta z}$,
where the work done by gas pressure is taken into account. 
The net energy liberation rate is $\Delta E_{\rm vis} \sim -\eta
\Omega_{\rm K} r P_{\theta z} \sim \eta^{2} v_{\rm K}^{2} \Omega_{\rm K}
T_{s} \Sigma_{d} / f_{\rm mid}$ (Equation (T18)), where we used $\nu
\sim h_{d}^{2} T_{s} \Omega_{\rm K}$ and $h_d = h_g \Sigma_d / (\Sigma_g
f_{\rm mid})$ (Equations (T16) and (T6)). The energy dissipation rate is
$\Delta E_{\rm turb}\sim\Sigma_{\rm layer}u_{\rm eddy}^{2}/\tau_{\rm eddy}\sim\alpha h_{g}
^{2}\Omega_{\rm K}^{3}\Sigma_{d}$, where $\Sigma_{\rm layer}\sim\Sigma_{d}$ is
dominated by the dust. Equating $\Delta E_{\rm turb}\sim C_{\rm eff}\Delta E_{\rm vis}$
gives
\begin{equation}
\alpha\sim(C_{\rm eff}\tilde{\eta}Z^{-1})^{2}T_{s} ~,
\label{eq:aplha-apx2}
\end{equation}
where we use $f_{\rm mid} \approx Z \sqrt{T_{s}/\alpha}$ (Equation
(T6)). Expressions (\ref{eq:aplha-apx1}) and (\ref{eq:aplha-apx2}) are
the same as Equation (T28), except for numerical factors and the
adoption of a slightly steeper power-law index ($\delta=1$ in Equation
(\ref{eq:aplha-apx2}) instead of $\delta=0.94$ in Equation (T28)) to
simplify analytical expressions. Connecting these expressions gives
\begin{equation}
\alpha\approx\left[  \left(  C_{1}C_{\rm eff}\tilde{\eta}Z\right)  ^{-\frac{2}{3}
}+\left(  C_{2}C_{\rm eff}\tilde{\eta}Z^{-1}\right)  ^{-2}\right]  ^{-1}
T_{s}~,\label{eq:alpha-apx3}
\end{equation}
where the numerical factors $C_{1}=1.0$ and $C_{2}=1.6$ are set to fit the
numerical result of T12. In Figure \ref{fig:alp-app}(a), the approximate
expression (\ref{eq:alpha-apx3}) is compared to the numerical result of T12.
The approximation overestimates $\alpha$ around $Z=\sqrt{C_{\rm eff}\tilde{\eta}}$
and underestimates it for large $Z$. However, the error is less than 2 for
$10^{-4}<Z<10^{-1}$, and the overall behavior appears to be acceptable.

The midplane dust-to-gas ratio $f_{\rm mid}=Z h_{g}/h_{d}\approx Z\sqrt
{T_{s}/\alpha}$ is, from Equations (\ref{eq:aplha-apx1}) and
(\ref{eq:aplha-apx2}),  $f_{\rm mid}\sim\lbrack Z^{2}/(C_{\rm eff}\tilde{\eta}
)]^{1/3}$ for $f_{\rm mid}\la1$, and $f_{\rm mid}\sim Z^{2}/(C_{\rm eff}\tilde{\eta})$ for
$f_{\rm mid}\ga1$. Connecting these two expressions gives an approximate
formula for $f_{\rm mid}$ as
\begin{equation}
f_{\rm mid}\approx\left(  \frac{Z^{2}}{C_{1}C_{\rm eff}\tilde{\eta}}\right)
^{\frac{1}{3}}+\left(  \frac{Z^{2}}{C_{2}C_{\rm eff}\tilde{\eta}}\right)
~.\label{eq:fmid2}
\end{equation}
From the above Equation (\ref{eq:fmid2}), it is seen that $f_{\rm mid}\approx1$ when $Z\approx (C_{\rm eff}\tilde{\eta})^{1/2}$.

The drift velocity of the dust in the limit of $f_{\rm mid}\ll1$ and $T_{s}\ll1$
is given in Equation (\ref{eq:vdr-ind}). In the limit of $f_{\rm mid}\gg1$,
$v_{d,r}\sim r P_{\theta z}/(v_{\rm K}\Sigma_{d})\sim \eta v_{\rm K}
T_{s} /f_{\rm mid}$ (Equation (T51)). Connecting these limits, we estimate
\begin{equation}
v_{d,r}\approx -\frac{2T_{s}\eta v_{\rm K}}{f_{\rm mid}+1}~,\label{eq:vdr}
\end{equation}
which gives the dust drift timescale given in Equation (\ref{eq:Tdrift}). This
approximate expression is compared with the numerical calculation by T12 in
Figure \ref{fig:alp-app}(b), which shows that the above expression provides a
good approximation. Note that $v_{d,r}\propto f_{\rm mid}^{-1}$ for $f_{\rm mid}\gg1$.
Equation (2.11) of Nakagawa et al. (1986) shows that, due to individual drag,
the radial velocity of the dust particles at the midplane is proportional to
$f_{\rm mid}^{-2}$. However, the average velocity of the dust layer,
$\int\rho_{d}v_{d,r} dz /\Sigma_{d}$, is approximately proportional to $f_{\rm mid}^{-1}$, provided that
the dust density profile $\rho_{d}(z)$ is Gaussian. The drift velocity due to
collective drag is also proportional to $f_{\rm mid}^{-1}$, as shown in Equation
(\ref{eq:vdr}). Thus, for $f_{\rm mid}>1$, the radial drift time derived from the
particle velocity at the midplane ($T_{\rm drift}\propto f_{\rm mid}^{2}$), which is
adopted in Y11, SC11, and MKI12, overestimates the average drift time
($T_{\rm drift} \propto f_{\rm mid}$), resulting in a less strict
condition for $T_{\rm grow}<T_{\rm drift}$ than reality. 


\end{document}